\documentclass[12pt]{article}
\usepackage{graphicx,a4} 
\usepackage{amssymb}
\usepackage{latexsym}
\begin{document}
\title{There is no explosion risk associated with superfluid Helium
    in the LHC cooling system}
\author{Malcolm Fairbairn and Bob McElrath\\
CERN theory group, Geneva 23, CH 1211, Switzerland}

\date{September 2008}

\maketitle
\begin{abstract}
We evaluate speculation about the possibility of a dangerous release of
energy within the liquid Helium of the Large Hadron Collider (LHC)
cryogenic system due to the occurrence of a ``Bose-Nova''.  Bose-Novae
are radial bursts of rapidly moving atoms which can occur when a Bose-Einstein
Condensate (BEC) undergoes a collapse due the interatomic potential being deliberately made attractive using a magnetic field close to the Feshbach resonance.
Liquid ${}^4$He has a monatomic structure with s-wave electrons, zero
nuclear spin, no hyperfine splitting, and as a consequence no Feshbach
resonance which would allow one to change its normally repulsive interactions to be
attractive.  Because of this, a Bose-Nova style collapse of ${}^4$He
is impossible.  Additional speculations concerning cold fusion during
these events are easily dismissed using the usual arguments about the
Coulomb barrier at low temperatures, and are not needed to explain the
Bose-Einstein condensate Bose-Nova phenomenon.  We conclude that that
there is no physics whatsoever which suggests that Helium could undergo 
any kind of unforeseen catastrophic explosion.
\end{abstract}

It has been suggested that there may be some danger associated with the
liquid Helium cryogenic system of the LHC.  The issue
under discussion is whether or not some kind of explosive event related
to the phenomenon known as a ``Bose-nova'' could occur within the liquid
Helium, as superfluid liquid ${}^4$He is a type of Bose condensate.  There
have also been suggestions that Bose-nova events are actually signals
of a type of cold nuclear fusion~\cite{claim1,claim2}.

Liquid Helium is a substance with a long history of usage, and
absolutely no history of spontaneous explosions.  The only hazard listed
by the U.S.  Occupational Safety and Health Administration is that it is
a ``simple asphyxiant''.  It was first liquefied in 1908, and is
routinely used all over the world by physics departments in condensed
matter experiments.  Furthermore, it has been used in high energy physics as a target in bubble chambers so the exercise of a beam impinging on
liquid Helium is one that has been repeated millions of times without
incident.  One recent example of the use of liquid Helium as a target is
at the Japanese RIKEN Laboratory~\cite{RIKEN}.

A Bose-Einstein condensate (BEC) can form when the de Broglie
wavelengths of individual particles become larger than their typical
separation, and the atoms condense into an entity exhibiting collective
dynamics~\cite{bose}.  Since the first production of a BEC in the mid
1990s~\cite{anderson95}, considerable progress has been made in
controlling their properties by manipulating the inter-atomic forces
between the atoms making up the condensates using magnetic fields.  In
particular, magnetic fields close to a Feshbach resonance of the atoms
in question can lead to large changes in the amount of
repulsion or attraction between atoms~\cite{inouye98}.

A Feshbach resonance occurs when the kinetic energy associated with the
scattering state between two atoms is degenerate with the energy of a
molecular bound state of those atoms.  Magnetic field strengths can be
tuned so that this degeneracy occurs in atomic systems where the bound
and unbound states possess different Zeeman shifts~\cite{herbig03}.
Tuning the interatomic interaction from repulsive to attractive therefore
requires a Feshbach resonance, which only occurs if the atom has
hyperfine structure.

In 2001, researchers used the Feshbach resonance to control the
interatomic attraction between atoms in a Bose-Einstein condensate of
Rubidium atoms and were able to induce an implosion of the condensate
resulting in atoms rapidly leaving the conglomeration, dubbed a
``Bose-nova'' by the authors~\cite{dynamics,Claussen01,Claussen02}.
Further studies investigated the dynamics of the explosion as a function
of the speed of the variation of the magnetic field.  Similar collapses
also occur in Bose-Einstein condensates formed from Lithium atoms where
the interatomic force is naturally attractive at zero magnetic field
strength~\cite{sackett99}.

The origin of the burst is well understood as being due to the release
of kinetic energy from local spikes in the atomic density that form
during the collapse~\cite{PhysRevA.65.033624,PhysRevA.66.011602}. An
additional component of atoms released during a collapse arises from
inelastic collisions between molecules formed at the Feshbach resonance
and other atoms and molecules, which becomes more important as the
density increases in a collapse~\cite{threebody1,threebody2,threebody3}.
There are no nuclear reactions involved in this generally accepted
explanation.

The collapse phenomenon is not relevant for the case of ${}^4$He.
${}^4$He has no hyperfine structure since the electrons are all in
s-wave orbits and there are no un-paired electrons, and furthermore it
has zero nuclear spin.  It is by tuning the hyperfine splitting that one
can change the properties of atoms with magnetic fields, so the
configuration of electrons and zero nuclear spin in ${}^4$He means that
there is no Zeeman effect and therefore no Feshbach resonance.  The
${}^4$He scattering length remains positive in all magnetic field
strengths, whereas a negative scattering length would be required to
create a Bose-Nova.

The release of chemical binding energy associated with the formation of
molecules in 3-body interactions is not possible due to the chemical
inactivity of ${}^4$He.  Bound states of ${}^4$He, held together by the
weak Van der Waals forces do exist,~\cite{toennies,luo} but the binding
energy which could be released in the formation of such states would
correspond to less than a thousandth of the thermal energy of the Helium
at 1.9 K~\cite{andersontraynor}.

Even if something unknown to science caused Helium to collapse or to
form molecules, this would simply heat the Helium until it was no longer
superfluid.  The LHC Helium system is specifically designed to dissipate
heat.  The LHC magnets and cryogenics are well prepared for a similar
situation in which the magnet spontaneously becomes non-superconducting
(quenching), releasing the stored magnetic field energy as heat.  Liquid
${}^4$He at 1.9K has very high specific heat and thermal conductivity
(around $10^5$ and $10^4$ times that of copper respectively) so that any
such heat energy will quickly be dissipated.  An input of energy from
the beam into the liquid Helium coolant would heat the Helium, obviating
the possibility of any condensed state forming as would be required to
instigate a Bose-Nova.  Magnets in the Tevatron accelerator at Fermilab
in the USA are superconducting and also employ liquid Helium.  These
magnets have been operating for more than a decade without problems, as
have many other superconducting accelerators.

Unlike astrophysical Super-Novae, the end result of a Bose-Nova cannot
be a black hole.  Black holes with masses below the observed Planck scale have been proposed as being
possible in theories containing extra
dimensions~\cite{Giddings:2001bu,Dimopoulos:2001hw}.  The safety
implications of these kind of black holes was thoroughly considered in
Ref.~\cite{Ellis:2008hg}.  To be consistent with the non-observation of
such objects, the fundamental Planck scale must be at or above the TeV
energy scale.  The energy of the Bose-Nova collapse is too small by at
least 14 orders of magnitude to create a black hole even in these
speculative low-Planck scale theories.

Finally, we address the speculation that the Bose-nova events are
actually signals of a type of cold nuclear fusion.  Fusion at low
temperatures has been strongly refuted by the physics community due to
the Coulomb repulsion of charged nuclei \cite{coldfusion}.  These
arguments apply equally well to any phenomena occurring in liquid Helium
or the ${}^{85}$Rb used in the original Bose-Nova experiment.  There
exists no possible sustainable reaction in any of the LHC components (none of the
LHC beam-line components are fissile).  There is simply no way that a
beam loss event could ``ignite'' the Helium.

Even if some fantastic physics in violation of Quantum Mechanics somehow
enabled Helium molecules to undergo a Bose-nova-style collapse resulting
in nuclear reactions, Helium has no energetically allowed nuclear
products in any two-body fusion or exothermic scattering mode.  The only
nuclear reaction pure Helium can undergo is three body, 3 ${}^4 {\rm
He}\to {}^{12}$C, a process which is proportional to the cube of the
density, and only occurs within the dense cores of old, massive stars.  It
cannot undergo sustained nuclear burning at the pressures and densities
used at CERN.

In conclusion, there is no science whatsoever, speculative or
otherwise, which suggests that anything dangerous could happen in the
Helium cooling system of the LHC.  Experience in a hundred years of use
in all kinds of physics experiments, as well as in high energy particle
physics, indicate beyond any reasonable doubt that Helium is safe and
cannot undergo any kind of unforeseen catastrophic explosion.

\section*{Acknowledgements}
We are very grateful to Elizabeth Donley of JILA and Mark Lee of University College London for looking over an earlier copy of this manuscript and John Ellis and Michaelangelo Mangano for additional comments.

\end{document}